\documentclass[letterpaper,10pt,conference]{ieeeconf}
\usepackage{amsmath,amsfonts}
\usepackage{algorithmic}
\usepackage{algorithm}
\usepackage{array}
\usepackage[caption=false,font=normalsize,labelfont=sf,textfont=sf]{subfig}
\usepackage{textcomp}
\usepackage{stfloats}
\usepackage{url}
\usepackage{verbatim}
\usepackage{graphicx}
\usepackage{color}
\usepackage{cite}
\usepackage{placeins}
\usepackage{needspace}
\IEEEoverridecommandlockouts
\title{A Co-Located MMG--FES Wearable Interface for Action-Updated State-Gated Stimulation}

\newif\ifanonymous
\anonymousfalse % true: double-anonymous review version; false: named version

\ifanonymous
\author{Anonymous Authors}
\else
\author{
Cong Liu$^{1}$, Xiao Jin$^{2}$, and Zhenhua Yu$^{1,2,*}$\\
\thanks{This work has been submitted to the IEEE for possible publication. Copyright may be transferred without notice, after which this version may no longer be accessible.}%
\thanks{*Corresponding author: Zhenhua Yu}%
\thanks{This work was partly supported by Capital Equipment funding of University of Aberdeen.}%
\thanks{$^{1}$Department of Computer Science, University of Aberdeen, AB24 3UE, Aberdeen, UK.
       }%
\thanks{$^{2}$Dyson School of Design Engineering, Imperial College London, SW7 2DB, UK.
        {\tt\small e-mail: zhenhua.yu@abdn.ac.uk}}%
}
\fi
\begin{document}
\bstctlcite{IEEEexample:BSTcontrol}
\maketitle
\thispagestyle{empty}
\pagestyle{empty}

\begin{abstract}
Functional electrical stimulation (FES) assists movement, but fixed schedules cannot track how a repeatedly activated muscle changes, and surface electromyography (EMG) near the stimulation electrodes is often obscured by artifacts. Prior work recorded mechanomyography (MMG) passively during closed-loop FES cycling and estimated FES-induced fatigue or predicted fatigue for FES control, without gating stimulation with stimulated-site MMG. In this study, we present VG-FES, a wearable module co-locating microphone-based MMG, programmable charge-balanced biphasic stimulation, and embedded state-gated control at one on-skin thigh site. After each action, we compare rectus femoris (RF) mean power frequency (MPF) with an early-session baseline of the same direction, giving a relative state. A threshold on its window means enables fixed-duration bursts separated by rest periods. During onboard stimulation, MMG remained recordable, whereas near-site EMG was artifact-dominated. Under prolonged external stimulation, cycle-wise MMG tracked an independent local-pressure reference (Spearman \(\rho=0.968\)). In leave-one-session-out evaluation on 2284 squat actions (11 sessions, eight participants), MMG-based action segmentation reached an event-detection F1 of 0.923 and direction classification a balanced accuracy of 0.925. Online, over 15.69~min of self-paced squats, one participant received four bursts while MMG acquisition and action monitoring continued. Mechanical feedback from the stimulated site offers wearable FES and assistive robotics a per-action basis for state-gated stimulation.
\end{abstract}

\begin{keywords}
Wearable Robotics; Human Factors and Human-in-the-Loop; Physically Assistive Devices; Mechanomyography; Functional Electrical Stimulation.
\end{keywords}

\section{Introduction}
\label{introduction}
Functional electrical stimulation (FES) activates skeletal muscle to assist movement and support rehabilitation after neurological injury~\cite{marquez2020fesreview}. A systematic review found that neuromuscular electrical stimulation (NMES) superimposed on voluntary contraction gave strength gains at least equal to voluntary training alone, with protocol-dependent results~\cite{borzuola2023superimposing}. Recent FES work includes textile-embedded sleeves~\cite{garnier2024novel}, networked bilateral stimulators~\cite{xu2024closed}, and hybrid FES--exoskeleton planning~\cite{mahan2025hybridplanning}. Muscle fatigue reduces force-generating capacity during repeated activation~\cite{enoka2008fatigue}, so the muscle's state changes over an exercise session. Model-based fatigue estimates have informed predictive knee-exoskeleton assistance~\cite{zhang2026fatigueaware}, and adaptive myoelectric control tracks changes in the user's motor strategy~\cite{peng2024adaptivecontrol}. However, FES is still commonly delivered on a preset schedule, while local muscle feedback could enable or suspend stimulation as exercise progresses. Stimulation enabled by such feedback cues the wearer that fatigue-related change has accumulated, and, at assistive intensities, it is also intended to supplement voluntary force output~\cite{borzuola2023superimposing}.

Electromyography (EMG) is widely used for muscle monitoring and has been combined with FES for lower-limb movement and fatigue-state recognition~\cite{zhang2024recognition,bala2024meffnet}. However, stimulation artifacts can obscure EMG recorded near the stimulated site, so acquisition relies on dedicated front-ends with real-time adaptive filtering~\cite{wang2021artifact,zhao2025stimemg}. Inertial sensing, force myography and A-mode ultrasound provide complementary information for FES feedback. Inertial sensors yield limb kinematics and task phase~\cite{schauer2017sensing}, and force myography reports surface-pressure changes produced by muscle deformation~\cite{xiao2019fmg}. A-mode ultrasound has measured stimulation-induced deformation that correlates with fatigue~\cite{alvarez2025stimulationinduced}. The mechanical vibration of the contracting muscle offers a further route for feedback at the stimulated site.

Mechanomyography (MMG) records this mechanical signal of contraction, the low-frequency oscillations generated by active fibers~\cite{orizio1993}. A review describes its use for muscle-function assessment and extraction of fatigue-related parameters~\cite{islam2013mmg}. Studies used MMG to estimate FES-induced neuromuscular fatigue~\cite{naeem2019mmgcycling,krueger2020mmgfatigue,11039026} and to predict fatigue for FES control~\cite{johnston2025mechanomyography}. Mean power frequency (MPF) has been used to define state labels during repetitive squat-lifting~\cite{wang2026mmgpilot}.

Mechanical muscle sensing has also been used in FES and robotic assistance. MMG was recorded in a closed-loop FES cycling system without informing the controller~\cite{woods2018cycling}. MMG and EMG have been combined for joint-torque modeling during exoskeleton interaction~\cite{caulcrick2021torque}. Capacitive microphone sensors have supported predictive and adaptive exoskeleton assistance~\cite{jin2026cmsassistance}. An MMG-informed glove triggers negative-pressure stiffening from recognized tremor without stimulating the sensed muscle~\cite{li2026mmgglove}. Co-located electrophysiological and mechanical sensing has been realized at one site~\cite{wang2026colocatedsensor}, and an FES electrode permits ultrasound imaging beneath it~\cite{moon2024ultrasoundelectrode}. However, the FES cycling controller did not use MMG. The robotic systems applied mechanical muscle sensing to an exoskeleton or a glove rather than to stimulation. Neither the fatigue estimates and predictions nor the co-located sensors were used to gate stimulation. The task here is to maintain MMG acquisition at the stimulated muscle while using an action-updated relative state to gate that stimulation within one wearable module.

This study examines the feasibility of such a module, which raises three challenges. First, MMG quality depends on sensor type and sensor--skin coupling~\cite{silva2003mmgnoise,posatskiy2012motion,castillo2021mmgforce}, and the sensor must share the thigh site with the stimulation electrodes. Second, during repeated movement MMG varies with movement phase and direction, so MMG from different actions is comparable only within the same phase and direction. Prior work segmented MMG by squat phase using inertial data~\cite{woodward2019segmenting}, and thus relied on an additional inertial sensor. Third, no absolute fatigue reference is available during voluntary movement, and single-action estimates fluctuate. The state must therefore be relative to the ongoing session, and the gating decision must tolerate this fluctuation.

To address these challenges, this study develops VG-FES, a compact module that attaches to the thigh and combines MMG sensing, programmable charge-balanced biphasic stimulation, and embedded processing. For the first, an air-filled conical structure couples muscle vibration from the skin to a MEMS microphone, and the stimulation electrodes share the same site. For the second, action segmentation and movement-direction classification use MMG itself, so no inertial sensor is added, aligned, or calibrated. Comparing each action with the early-session baseline of its direction avoids a per-user calibration session. For the third, the decline of rectus femoris MPF relative to this baseline forms a relative state. A window-mean gate with a single threshold enables fixed-duration onboard stimulation bursts separated by rest periods. Fig.~\ref{fig:experiment_description} shows the thigh placement of VG-FES and the repeated-squat protocol.

The main contributions of this study are as follows:

\begin{enumerate}
\setlength{\itemsep}{0pt}
\setlength{\parsep}{0pt}

\item We developed a vertically stacked module that attaches to the thigh and co-locates MMG sensing, programmable charge-balanced biphasic stimulation, and embedded processing.

\item We showed that MMG remained recordable during onboard stimulation while near-site EMG was artifact-dominated, and that cycle-wise MMG followed the local-pressure trend under separate external stimulation.

\item We implemented a closed-loop controller that gates fixed-duration stimulation bursts between rest periods while MMG acquisition continues, and demonstrated it in an online closed-loop session. The gating variable is an MMG state that is updated after each action relative to an early-session baseline. MMG-based action segmentation and direction classification need no additional inertial sensor, and early-session baselines avoid per-user calibration sessions.
\end{enumerate}

\section{Co-Located Wearable MMG-FES Feedback Platform}
\label{design}

\subsection{Platform Architecture and Co-Located Interface}
\begin{figure*}[!t]
\centering
\includegraphics[width=0.9\textwidth]{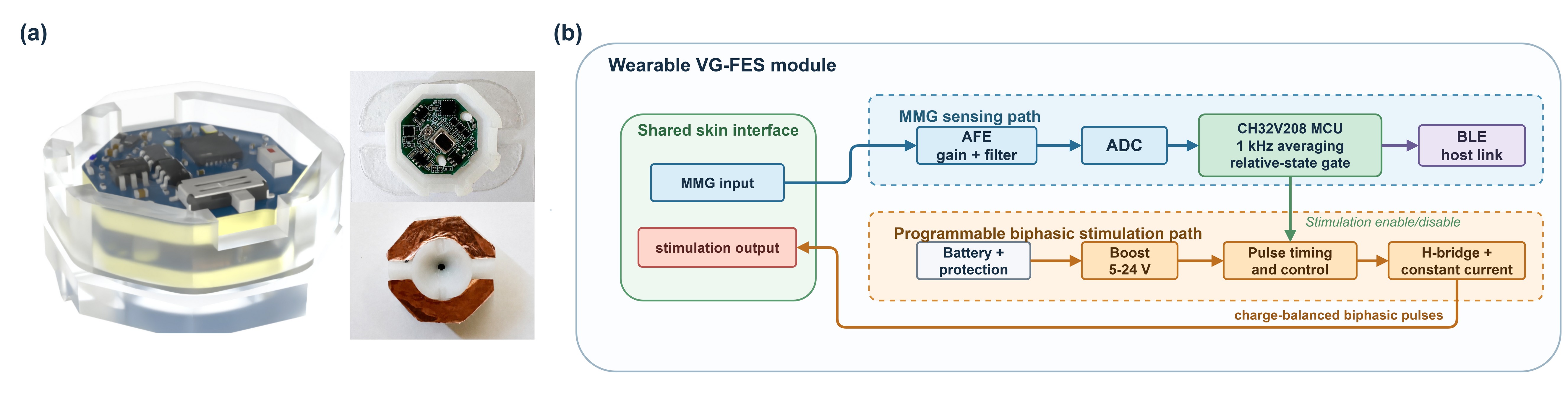}
\caption{Wearable VG-FES platform. (a) Top-side and skin-contact views of the co-located module. (b) Functional architecture of MMG sensing, biphasic stimulation, embedded processing, and BLE communication. Action detection on the host and relative-state estimation on the module feed the gate for fixed-parameter stimulation.}
\label{fig:overview_double_panel}
\end{figure*}

VG-FES co-locates MMG sensing and electrical stimulation in one vertically stacked wearable module that attaches to the skin of the thigh. From the skin upward, the stack comprises a pair of adhesive stimulation electrodes, an air-filled conical acoustic-coupling structure, and two printed circuit boards (PCBs). Silicone potted around and between the PCBs provides mechanical support. The acoustic-coupling structure, a MEMS microphone, and the analog front-end (AFE) form the MMG acquisition path. A flexible ribbon cable connects the upper and lower PCBs, and the octagonal housing, measuring approximately 35~$\times$~34~$\times$~14~mm, accommodates the stacked PCBs and coupling structure. Fig.~\ref{fig:overview_double_panel} shows the assembled module and its MMG acquisition, stimulation, processing, and host-communication paths, and Fig.~\ref{fig:hardware_coupling_placement}(a) shows the octagonal PCBs before panel separation.

\subsection{Acoustic Coupling and MMG Signal Acquisition}
\begin{figure*}[!b]
\centering
\includegraphics[width=\textwidth]{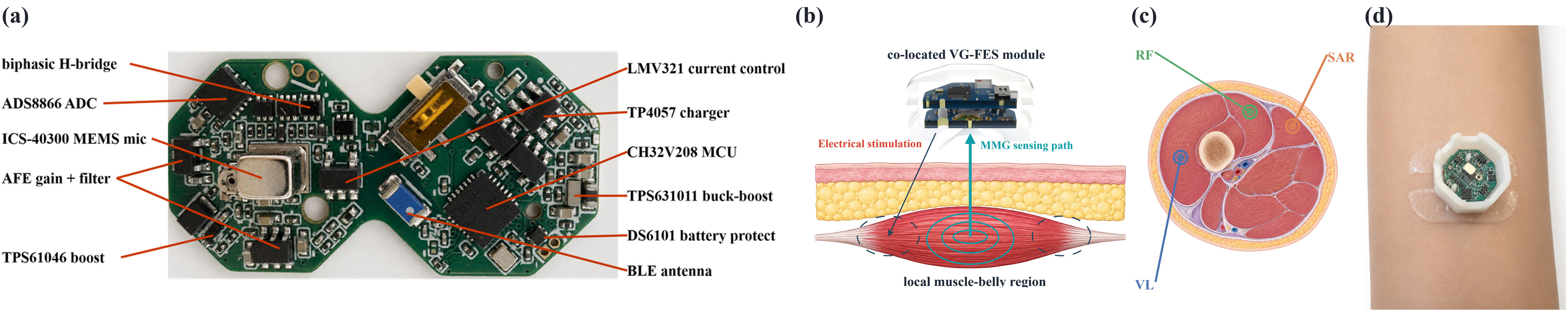}
\caption{VG-FES hardware, tissue coupling, and thigh sensing-site placement. (a) Annotated PCB panel before separation into the octagonal upper and lower boards. (b) Sectional illustration of the tissue-to-module vibration-transfer path. (c) Cross-sectional thigh schematic showing the sartorius (SAR), rectus femoris (RF), and vastus lateralis (VL). (d) Representative on-skin mounting of the co-located module.}
\label{fig:hardware_coupling_placement}
\end{figure*}

The conical acoustic-coupling structure sits above the electrode layer and has a small opening at its skin-facing apex. Low-frequency vibrations generated by muscle contraction pass through this opening, the air inside the cone, and the acoustic port in the lower PCB to reach the MEMS microphone. Fig.~\ref{fig:hardware_coupling_placement}(b) shows this local tissue-to-sensor path. The recorded MMG signal can also be affected by environmental and movement-related noise~\cite{silva2003mmgnoise,posatskiy2012motion}.

\begin{figure}[!t]\vspace*{4pt}
\centering
\includegraphics[width=\linewidth]{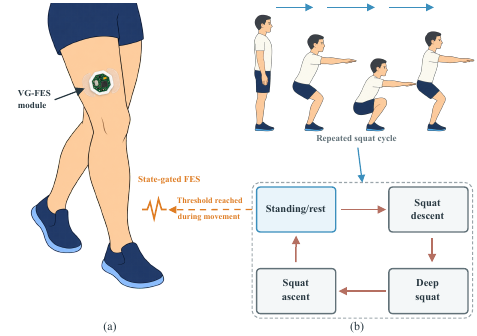}
\caption{VG-FES placement and closed-loop squat protocol. (a) On-thigh placement of the co-located MMG--FES module. (b) Repeated squat cycle and four-state action representation used for action tracking and relative-state updating. A single-threshold gate triggers fixed-duration stimulation bursts separated by rest periods.}
\label{fig:experiment_description}
\end{figure}

The lower PCB integrates an ICS-40300 analog MEMS microphone and a mid-supply-referenced AFE. The AFE AC-couples and biases the microphone output and includes an active-gain stage with approximately 32--33~dB of midband gain. Its first-order high-pass and low-pass filters have corner frequencies near 0.1~Hz and 220~Hz, respectively. This passband covers the low-frequency MMG content targeted in this study~\cite{orizio1993,islam2013mmg}. An ADS8866 analog-to-digital converter (ADC) digitizes the conditioned signal at 25~kHz. A CH32V208 microcontroller on the upper PCB block-averages the samples to 1~kHz for transmission to the host in buffered Bluetooth Low Energy (BLE) packets.

\subsection{Programmable Biphasic Stimulation Electronics}
The lower PCB contains a driver for programmable, charge-balanced biphasic stimulation~\cite{merrill2005stimulation}, in which a boost converter provides adjustable compliance voltage and a full H-bridge reverses polarity. A closed-loop constant-current controller regulates the output current as skin--electrode impedance varies with sweat, motion, or minor electrode displacement. The driver supports 5--24~V compliance, 0--5~mA output, 1-$\mu$s pulse-timing resolution, and repetition frequencies from 1~Hz to below 10~kHz. Short silver-coated wires connect the stimulation output to the adhesive electrode pair. The pressure-reference experiment in Section~\ref{exp} used a separate external stimulator.

\subsection{Embedded Control, Wireless Communication, and Power Management}
In the evaluated closed loop, the CH32V208GBU6 microcontroller updated the relative state after each completed action. It applied a single-threshold gate to 30-s window means of this state and timed stimulation as fixed-duration bursts separated by rest periods. The stimulation circuitry allows configuration of amplitude, pulse width, and frequency. MMG acquisition and action monitoring remained active during stimulation, with stimulation parameters held fixed.

During operation, the microcontroller streamed the 1~kHz MMG signal and the stimulation status to the host over BLE. The host logged and displayed the data, ran MMG-based action segmentation and movement-direction classification (Section~\ref{sec:ml_framework}), and computed the MPF of each completed action. Each MPF value was returned over BLE to the microcontroller, which updated the relative state and gated stimulation. Because the relative state changed once per completed action and the gate acted on 30-s window means, BLE latency was negligible on these time scales. The power-management circuitry of the upper PCB provides battery charging and protection and supplies regulated power under varying operating loads.

\section{Stimulation-Compatible MMG Sensing and On-Body Response Validation}
\label{exp}

The experiments, which involved repeated everyday lower-limb tasks and transcutaneous electrical stimulation in a controlled laboratory, were approved by \ifanonymous the institutional ethics committee (approval ID withheld for anonymous review)\else the University of Aberdeen Ethics Committee (ID: 8891100)\fi, and informed consent was obtained from all participants.

\subsection{MMG and EMG Recording During Concurrent Onboard Stimulation}
\begin{figure}[!t]\vspace*{4pt}
\centering
\includegraphics[width=\linewidth,trim=16pt 5pt 1pt 0pt,clip]{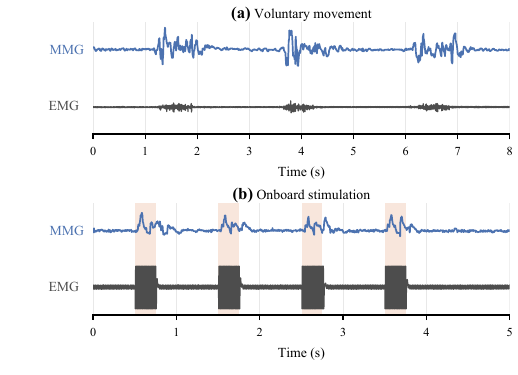}
\caption{Synchronized MMG and near-site EMG during voluntary movement and onboard stimulation. (a) Both channels show contraction-related activity during voluntary movement. (b) During four 250-ms stimulation intervals (shaded), EMG was dominated by stimulation artifacts, whereas MMG remained within range and retained temporal structure. Stimulation caused slight tremor without overt contraction.}
\label{fig:emg_artifact_example}
\end{figure}

In one participant, we recorded synchronized MMG and EMG over the rectus femoris (RF) on the anterior thigh. EMG electrodes were placed near the module on the same muscle (near-site EMG). Recordings covered voluntary movement and onboard stimulation [Fig.~\ref{fig:emg_artifact_example}]. The test assessed whether MMG remained recordable during stimulation at the same skin site. Onboard stimulation used the module's paired adhesive electrodes, which flank the central MMG contact on the skin-facing surface. The stimulation current was set to 2.5~mA, and the controller issued four 250-ms commands at approximately 1-s intervals.

Voluntary movement produced contraction-related responses in both MMG and EMG channels. During onboard stimulation, near-site EMG was dominated by stimulation artifacts that saturated the channel, whereas MMG remained within its recording range and retained visible temporal structure. This stimulation elicited slight tremor without overt muscle contraction. A separate external-stimulation experiment assessed whether cycle-wise MMG changes corresponded to the local mechanical response, using an independent pressure reference.

\begin{figure}[!htb]
\centering
\includegraphics[width=\linewidth]{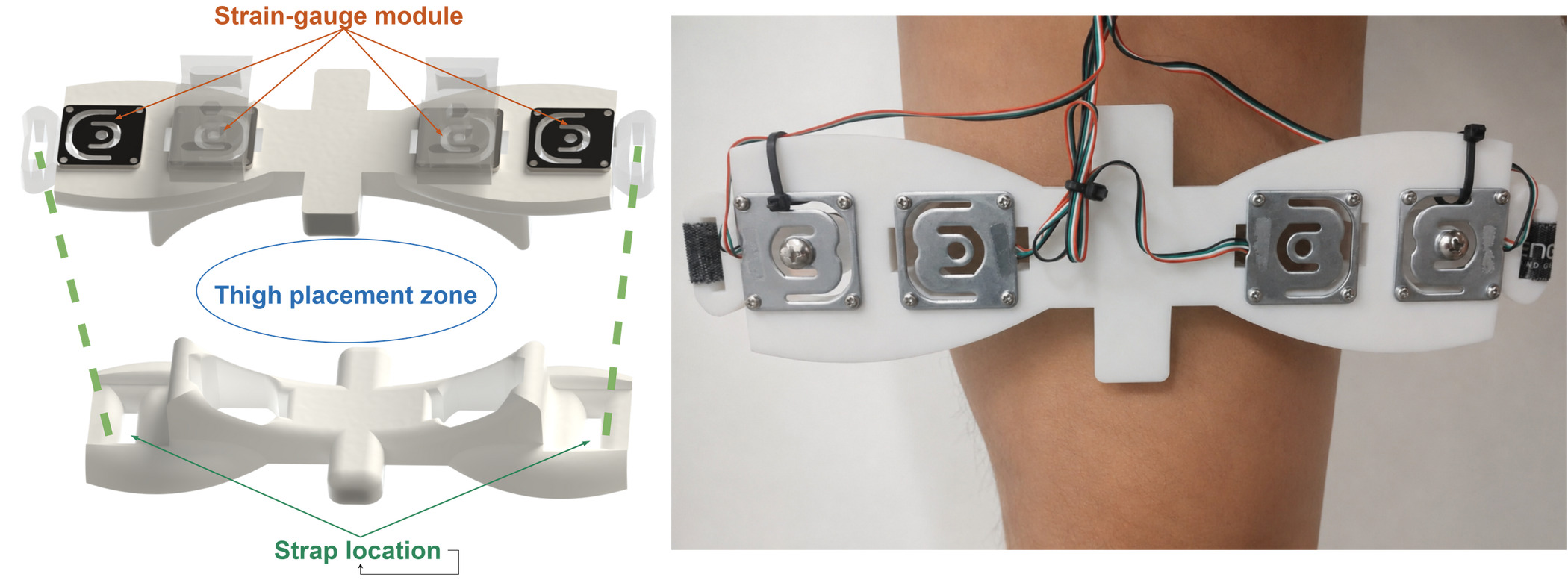}
\caption{Contact-pressure support fixture for the pressure-reference experiment. (a) Fixture geometry, strain-gauge modules, and thigh placement zone. (b) Worn configuration showing strapping and MMG-contact alignment over the target muscle.}
\label{fig:onbody_fixture_setup}
\end{figure}

\Needspace{7\baselineskip}
\subsection{Pressure-Referenced On-Body Evaluation Protocol}
\label{sec:pressure_protocol}

\subsubsection{Pressure Fixture and Muscle-Site Placement}
The thigh-strapped contact-pressure fixture in Fig.~\ref{fig:onbody_fixture_setup} carried strain-gauge modules, and local contact pressure was the average of their readings. The VG-FES module was placed under the distal end of the fixture's central cross [Fig.~\ref{fig:onbody_fixture_setup}(b)], which held its MMG contact over the selected muscle belly during repeated stimulation. Muscle contraction deforms the tissue and changes local contact pressure. These changes served as a relative reference for contraction magnitude~\cite{xiao2019fmg}. Before each run, strap tension was adjusted with the muscle relaxed to establish a stable, nonsaturated pressure baseline. This preload limited variation in initial normal force, which can affect MMG recording~\cite{castillo2021mmgforce}.

The three recording sites were over RF, the vastus lateralis (VL), and the sartorius (SAR) [Fig.~\ref{fig:hardware_coupling_placement}(c)]. RF and VL contribute to knee extension, whereas SAR spans the hip and knee and provides an anatomically distinct anteromedial site. Each site was selected where the muscle belly bulged visibly during contraction and was confirmed by palpation. These sites lay 12--15~cm above the patella on the anterior thigh for RF, 14--16~cm on the anterolateral thigh for VL, and 10--13~cm on the anteromedial thigh for SAR.

\begin{figure}[!t]\vspace*{4pt}
\centering
\includegraphics[width=0.9\linewidth]{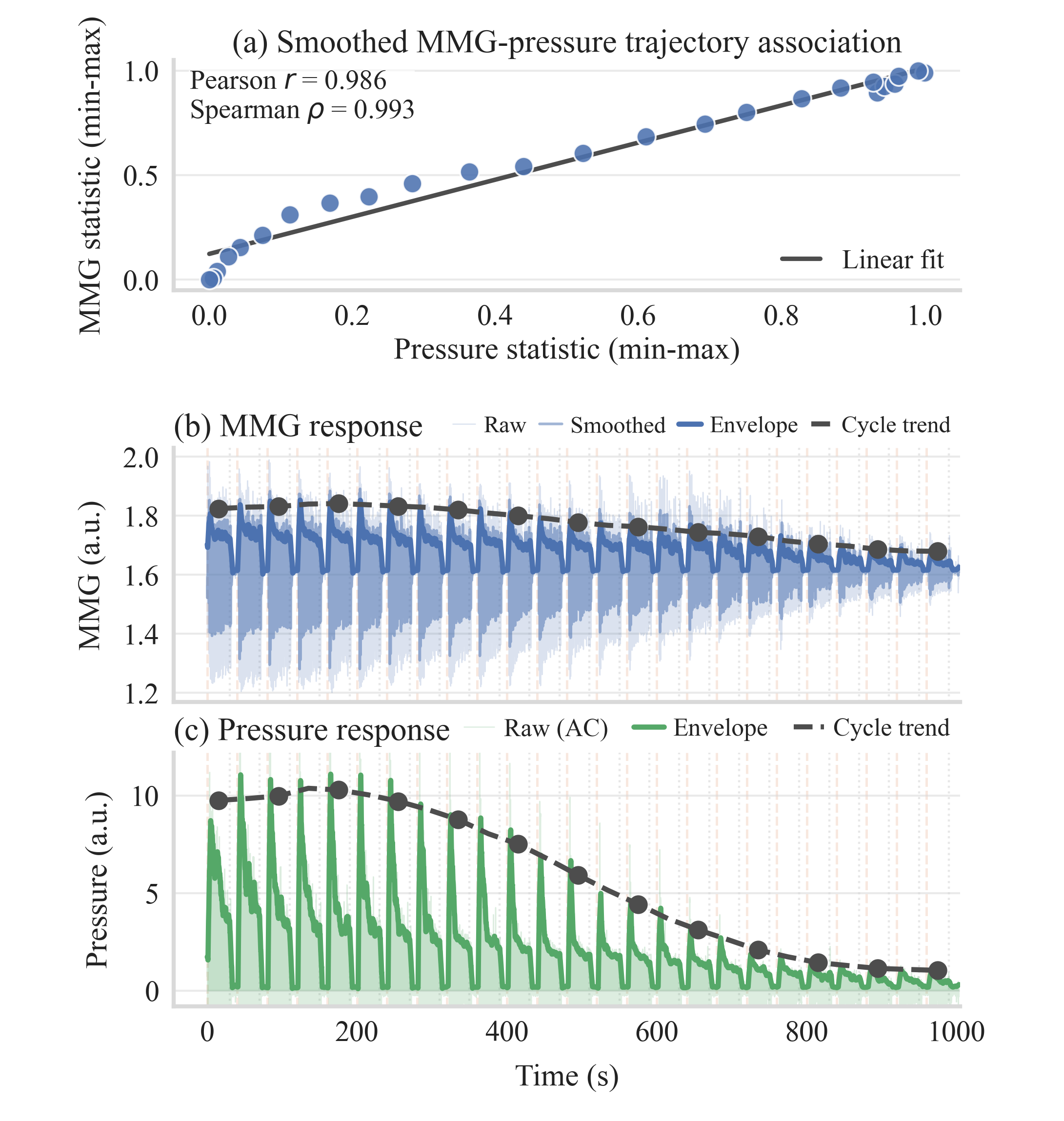}
\caption{Representative 1000-s RF-site response across 25 externally stimulated contraction--rest cycles. (a) Smoothed VG-FES MMG and local-pressure responses (Pearson \(r=0.986\); Spearman \(\rho=0.993\)), separately min--max normalized. (b) MMG envelope and cycle-response trajectory. (c) Pressure-reference response showing stronger buildup and sustained inter-cycle decline. Dashed and dotted lines indicate cycle onset and scheduled stimulation offset.}
\label{fig:fatigue_ems_example}
\end{figure}

\subsubsection{Repeated-Stimulation Protocol}
The pressure-reference experiment involved three participants. RF, VL, and SAR were tested in separate sessions in randomized order, with at least 24~h between repeated tests in the same participant. A commercial dual-channel NMES stimulator delivered 5--10~mA, with a compliance limit of approximately 30~V. Electrodes measuring approximately \(3.5~\times~5.7~\mathrm{cm}\) were placed longitudinally, with the MMG contact between the two stimulation electrodes. At each site, current was increased until a clear but tolerable contraction was visible or palpable and was then held constant.

Each 1000-s test comprised 25 cycles of 30-s stimulation and 10-s rest [Fig.~\ref{fig:fatigue_ems_example}], with electrode placement, stimulator settings, and fixture preload held constant. Within each \mbox{30-s} stimulation period, pulses were delivered at 100~Hz in 0.5-s bursts separated by 0.5-s off intervals. Each pulse comprised an approximately 500-$\mu$s active phase followed by an opposite-polarity, exponentially decaying recovery phase, with no programmed interphase interval.

\subsubsection{Cycle-Wise Signal Processing and Fatigue-Related Trend}
MMG and pressure recordings were aligned by their timestamps for offline analysis. Cycles were defined from the recording start using the prespecified 30-s stimulation and 10-s rest schedule. Centered rolling filters were used to construct envelopes for both signals. The MMG envelope retained its local baseline, whereas the pressure envelope was derived from baseline-subtracted pressure.

For each signal, the cycle response was defined as the 98th percentile of the processed envelope within its 30-s stimulation window. The MMG and pressure responses from the same cycle formed one pair. The resulting cycle-response sequences were smoothed with a centered five-cycle moving mean. Pearson and Spearman correlations were calculated between paired MMG and pressure responses from the same recording, both before and after this smoothing. The smoothed MMG and pressure trajectories were separately min--max normalized for visualization.

After the initial response buildup, we took a sustained relative decline in the smoothed local-pressure cycle responses as the fatigue-related trend for this protocol. This interpretation is consistent with the reduction in force-generating capacity associated with muscle fatigue~\cite{enoka2008fatigue}. Local pressure served as the reference, and we compared the smoothed MMG trajectory with this trend. We assessed changes across cycles because attenuation within individual stimulation windows could also reflect burst-onset adaptation or contact settling.

\subsection{Externally Stimulated MMG--Pressure Response Trend}

Fig.~\ref{fig:fatigue_ems_example} shows VG-FES MMG and the independently measured local-pressure response at the RF site in one participant during external stimulation. Before five-cycle smoothing, paired MMG and pressure cycle responses yielded Pearson \(r=0.974\) and Spearman \(\rho=0.968\). After smoothing, the corresponding coefficients were \(r=0.986\) and \(\rho=0.993\). Both trajectories showed an initial buildup followed by a sustained decline across cycles [Fig.~\ref{fig:fatigue_ems_example}(b), (c)], which is the fatigue-related trend defined in Section~\ref{sec:pressure_protocol}.

The coefficients describe similarity in the slow MMG and pressure trends within this recording. Pressure provided a higher-contrast mechanical reference. MMG was noisier but followed the same overall direction. For voluntary movement, the next section evaluates action segmentation and movement-direction classification as the basis for action-level MMG comparisons.

\section{MMG-Based Action Monitoring}
\label{sec:ml_framework}

\begin{figure*}[!t]\vspace*{4pt}
\centering
\includegraphics[width=0.95\textwidth]{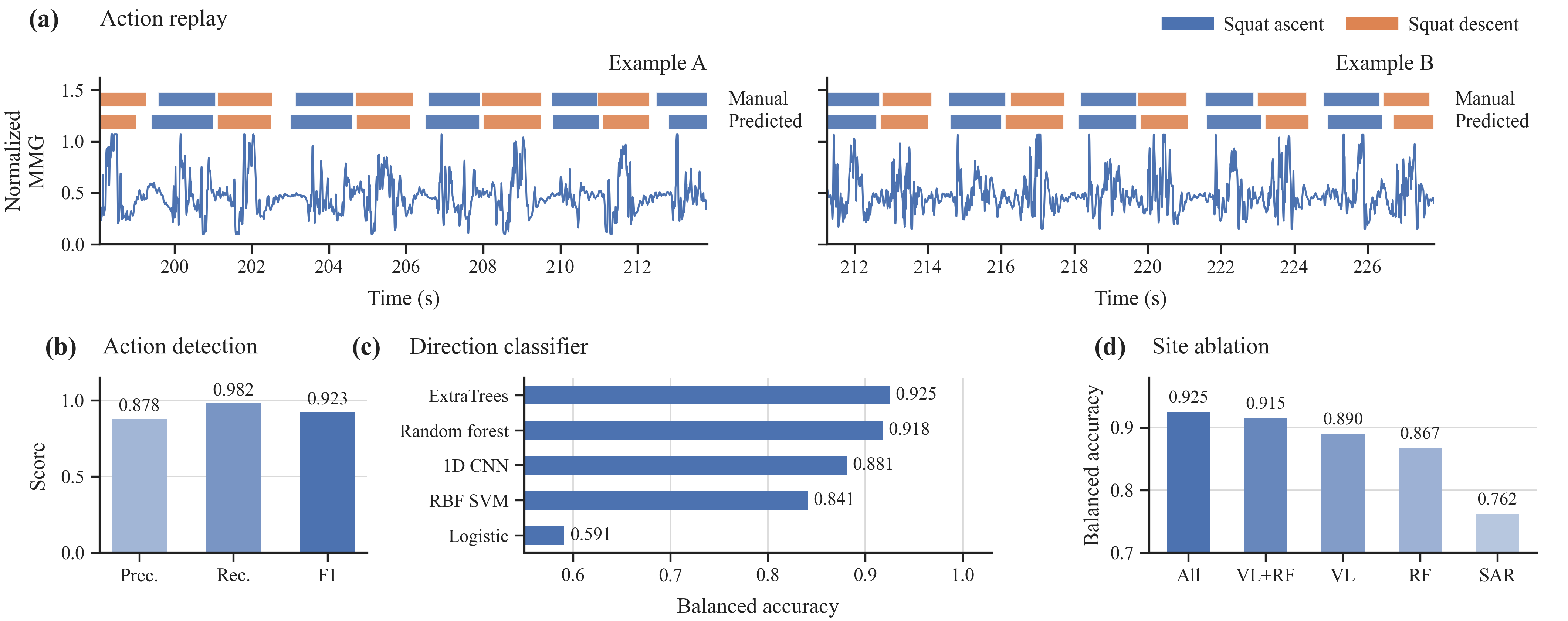}
\caption{MMG-based action monitoring. (a) Action intervals and model replay. (b) Action-event precision, recall, and F1. (c) Direction-classifier comparison using manual boundaries. (d) Channel and muscle-site ablation.}
\label{fig:action_protocol_model_results}
\end{figure*}

\begin{figure*}[!t]
\centering
\includegraphics[width=0.95\textwidth]{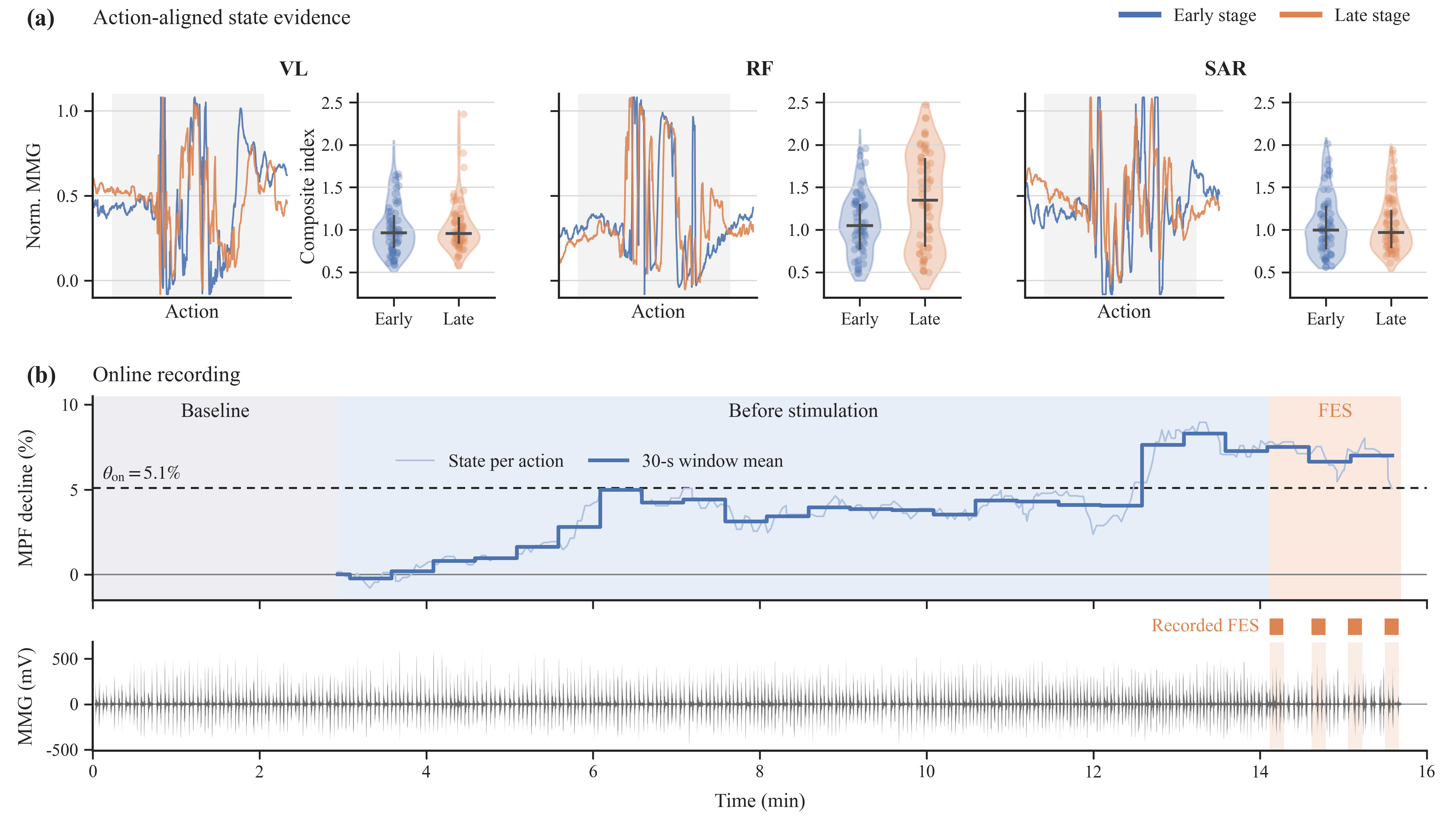}
\caption{Action-aligned MMG analysis and state-gated onboard stimulation. (a) Early- and late-stage MMG waveforms and pooled baseline-normalized composite-index distributions across three muscle sites. (b) Online closed-loop session. Top: per-action state \(d_b\) (light) with 30-s window means (dark step) and \(\theta_{\mathrm{on}}\) (dashed); shading marks baseline, pre-stimulation, and stimulation periods. Middle: bars mark the four recorded bursts. Bottom: band-pass-filtered RF MMG on a common time axis.}
\label{fig:action_fatigue_closed_loop}
\end{figure*}

\subsection{Repeated-Motion Dataset and Action Units}
The voluntary-movement dataset comprised repeated squat-descent and squat-ascent movements, with MMG recorded synchronously at RF, VL, and SAR using the placement and attachment shown in Fig.~\ref{fig:hardware_coupling_placement}(c,d). During each session, the participant performed self-paced alternation between upright standing and a deep-squat posture following the movement cycle illustrated in Fig.~\ref{fig:experiment_description}(b). Each squat descent or ascent was treated as one action. Action boundaries and movement directions were manually annotated and visually checked against the multichannel MMG signals. Unreliable segments at the ends of recordings were excluded from model development. After quality screening, 2284 complete actions from 11 sessions recorded from eight participants were retained: 1143 squat-ascent and 1141 squat-descent actions. Leave-one-session-out cross-validation held out one complete session for testing and used the remaining sessions for training. This kept actions from the same session together and assessed transfer across session-specific contact and movement-rhythm differences~\cite{roberts2017cv}.

\subsection{MMG-Based Action Segmentation and Movement-Direction Classification}
For action segmentation, a causal temporal convolutional network (TCN)~\cite{lea2017tcn} distinguished action from background using per-channel time-domain amplitude, variability, and slope statistics from 0.5-s MMG windows at 0.1-s intervals. Predicted action intervals were matched one-to-one to manual annotations using an intersection-over-union (IoU) threshold of 0.20. Precision, recall, and F1 were calculated separately for each held-out session and averaged across the 11 sessions.

{\widowpenalty=10000 For movement-direction classification, MMG signals were band-pass filtered at 5--100~Hz, and action waveforms were extracted using manually annotated boundaries. Each waveform was resampled to 256 points per channel and labeled as squat ascent or descent. Logistic regression, an RBF-kernel support vector machine (SVM), random forest, extremely randomized trees (ExtraTrees)~\cite{geurts2006extra}, and a compact one-dimensional convolutional neural network (CNN) baseline were compared. Using manual boundaries isolated the direction-classification assessment from action-segmentation errors.\par}

Across the 11 held-out sessions, the segmenter achieved a mean event-detection F1 score of 0.923 [Fig.~\ref{fig:action_protocol_model_results}(b)]. For the matched events, the predicted onset occurred \(0.278\pm0.542\) s after the manual onset, and the predicted offset occurred \(0.201\pm0.412\) s before the manual offset. Using all MMG channels and manually annotated action boundaries, ExtraTrees achieved accuracy, balanced accuracy, and macro F1 of 0.925 across 2284 held-out predictions (per-session balanced accuracy \(0.929\pm0.040\)) [Fig.~\ref{fig:action_protocol_model_results}(c)]. It correctly classified 1073 squat-ascent and 1039 squat-descent actions, with 172 errors. Among the compared classifiers, ExtraTrees gave the highest balanced accuracy, followed by random forest (0.918). Fig.~\ref{fig:action_protocol_model_results}(a) compares model outputs with manually annotated action intervals in a representative replay.

Channel ablation assessed how the MMG sensing sites contributed to movement-direction classification. Balanced accuracy was 0.925 with all channels, 0.915 with VL+RF (per-session \(0.916\pm0.066\)), 0.890 with VL alone, 0.867 with RF alone, and 0.762 with SAR alone [Fig.~\ref{fig:action_protocol_model_results}(d)]. The VL+RF pair gave higher balanced accuracy than either site alone and was 0.010 below the all-channel result. This pattern suggests that VL and RF provide complementary information for distinguishing squat ascent from descent.

\section{Relative-State Estimation and State-Gated Stimulation}
\label{sec:closed_loop}

Fig.~\ref{fig:action_fatigue_closed_loop} shows multifeature MMG analysis (a) and online state-gated stimulation (b). The online closed-loop session tested whether updates of an MMG-derived relative state (Section~\ref{sec:relative_state}) after each completed action could gate fixed-duration onboard stimulation bursts during voluntary movement.

\subsection{Action-Aligned MMG State Indicators}
MMG features were extracted from manually annotated action intervals at RF, VL, and SAR. The descriptive analysis in Fig.~\ref{fig:action_fatigue_closed_loop}(a) used action-site observations from the same voluntary-movement dataset. Each observation represented one action recorded at one muscle site. The features comprised envelope and root-mean-square (RMS) amplitude, median frequency (MF), mean power frequency (MPF), and bandwidth.

Within each session and muscle site, each feature was normalized to the median of its first five valid smoothed values. Early and late stages were defined as the first and final thirds of each session's action sequence. For each feature and muscle site, normalized observations were pooled by stage to obtain early- and late-stage medians. Percentage changes were calculated as \(100(m_{\mathrm{late}}/m_{\mathrm{early}}-1)\), where \(m_{\mathrm{early}}\) and \(m_{\mathrm{late}}\) are the corresponding early- and late-stage medians. A composite index combined baseline-relative increases in envelope and RMS amplitude with decreases in MF, MPF, and bandwidth~\cite{johnston2025mechanomyography}.

The early-to-late changes differed across muscle sites. At RF, the median normalized envelope and RMS increased by 12.5\% and 22.1\%, while the corresponding MPF and bandwidth medians decreased by 4.0\% and 10.3\%. VL showed decreases in both amplitude and spectral descriptors, whereas SAR showed little amplitude change and decreases in MPF and bandwidth. The median composite index increased at RF from 1.05 to 1.35, whereas VL (0.97 to 0.96) and SAR (1.00 to 0.97) remained near baseline [Fig.~\ref{fig:action_fatigue_closed_loop}(a)]. Among six faster sessions (about 13--19 squats per min), RF MPF decreased in five (5.7--8.8\%) and increased by 5.0\% in the sixth. In five slower sessions (about 6--9 squats per min), RF MPF did not decrease (+0.1 to +7.1\%). This difference is consistent with slower fatigue accumulation at a lower pace, although pace was uncontrolled. Nevertheless, after within-session normalization, the duration of individual actions was not associated with RF MPF across all sessions (Spearman \(\rho=-0.03\)).

\subsection{Relative-State Definition}
\label{sec:relative_state}
For each completed action \(b\), RF MPF was estimated from its central 0.8 s, which limited the effect of action-boundary errors (Section~\ref{sec:ml_framework}). Baselines and smoothing were computed separately for squat ascent and squat descent because RF MPF differed between the two directions. For each direction, the baseline \(\mathrm{MPF}_{\mathrm{base}}\) was the median MPF of the first 30 actions. The smoothed estimate \(\widetilde{\mathrm{MPF}}_b\) was the median MPF of the latest 30 actions available at the completion of action \(b\). The decrease \(D = 1 - \widetilde{\mathrm{MPF}}_b / \mathrm{MPF}_{\mathrm{base}}\) was computed for each direction, giving \(D^{\uparrow}\) for ascent and \(D^{\downarrow}\) for descent. The relative state was defined as the mean of the latest ascent and descent decreases:
\begin{equation}
d_b = \tfrac{1}{2}\left(D^{\uparrow} + D^{\downarrow}\right),
\end{equation}
and \(d_b\) was held at 0 until both directions had 30 actions for their baselines. Positive values indicate MPF below baseline, and negative values indicate MPF above baseline.

\subsection{State Updates and Stimulation Scheduling}
After each completed action, \(d_b\) was updated and the gate re-evaluated. The gate averaged \(d_b\) over consecutive 30-s windows and enabled stimulation when the means of three consecutive windows were all at or above \(\theta_{\mathrm{on}}\). Each opening delivered one 10-s burst, then closed automatically. A \mbox{15-s} rest period followed each burst, during which stimulation could not be re-enabled. The next completed action afterward re-enabled stimulation if the window condition still held. We set \(\theta_{\mathrm{on}} = 5.1\%\) at two times the expected fluctuation of the smoothed state, estimated from action-to-action MPF variability in recorded voluntary-movement sessions. Window averaging reduces this fluctuation, so applying \(\theta_{\mathrm{on}}\) to window means is conservative, and requiring three consecutive windows further limits spurious enabling.

We tested the controller online with one participant performing self-paced squats, module over RF, for 15.69~min. The host ran the Section~\ref{sec:ml_framework} action-segmentation and movement-direction models on MMG streamed from the module, then computed per-action MPF and returned it to the module. Stimulation consisted of charge-balanced biphasic pulses at 1.75~mA, delivered in 10-s bursts separated by 15-s rest periods. Discarding detections within 1.0~s of the previous one left 337 actions (about 11 squats per min). Per-action MPF, \(d_b\), and stimulation output were recorded on the MMG timeline.

Fig.~\ref{fig:action_fatigue_closed_loop}(b) shows the online closed-loop session. Stimulation was first enabled at 14.11~min, after 309 actions, when three consecutive 30-s window means of \(d_b\) crossed \(\theta_{\mathrm{on}}\). Four 10-s bursts were delivered, with onsets 30.2, 26.2, and 26.4~s apart. MMG acquisition and action detection continued throughout. Eleven actions were detected during stimulation.

\section{Discussion}
MMG recording remained active during state-gated stimulation. Earlier studies used MMG to estimate or predict fatigue~\cite{11039026,johnston2025mechanomyography}. Here, the action-updated relative state of RF determined stimulation delivery.

RF amplitude increased during voluntary movement but declined after an initial buildup under fixed external stimulation, and trends also differed among muscle sites. These differences show why MMG changes need to be interpreted in relation to the muscle and task. Movement and sensor coupling can also affect MMG measurements~\cite{posatskiy2012motion,castillo2021mmgforce}. To limit these effects, stimulation control compared RF MPF with a baseline from the first 30 actions in the same direction (Section~\ref{sec:relative_state}).

This study has several limitations. Fatigue was not measured independently during voluntary movement, so the observed RF changes cannot be attributed to fatigue alone. Direction classification used manually marked action boundaries, and the leave-one-session-out evaluation was organized by session rather than by participant. In addition, the online test comprised one session in one participant, and it did not evaluate whether the stimulation relieved fatigue or restored force.

\section{Conclusion}
This study examined a wearable FES module with co-located MMG sensing, programmable biphasic stimulation, and embedded processing. MMG remained recordable during onboard stimulation while near-site EMG was artifact-dominated, and cycle-wise MMG followed a local-pressure reference. Action segmentation achieved an F1 of 0.923 across 11 sessions comprising 2284 annotated actions, and direction classification reached a balanced accuracy of 0.925. In a 15.69~min online closed-loop session, the relative state was updated after each action. A gate with a single threshold on this state delivered four fixed-duration bursts while MMG acquisition continued.

Prior work recorded MMG in FES loops without informing control or estimated fatigue from MMG. Here, co-locating sensing with stimulation and gating on a per-action baseline-relative state yields a control variable at stimulation sites where stimulation artifacts obscure near-site EMG, providing a basis for state-gated wearable FES. Future work will evaluate end-to-end direction classification on model-segmented action boundaries, assess generalization with participant-wise cross-validation, and characterize muscle- and task-specific MMG features across additional muscle sites and movements. Future work will also extend the online closed-loop test to multiple participants and sessions, add an independent fatigue reference during voluntary movement, use force and perceived-fatigue measures to evaluate whether state-gated stimulation relieves fatigue and restores force, and integrate the module with FES-assisted exercise and wearable robotic assistance.

\bibliographystyle{IEEEtran}
\bibliography{bibs}
\end{document}